# Optimal upper bounds for non-negative parameters

Fyodor V. Tkachov
Institute for Nuclear Research of the Russian Academy of Sciences
Moscow 117312, Russian Federation
ftkachov@ms2.inr.ac.ru

**Abstract.** Using the techniques of [arXiv:0911.4271], upper bounds for a given confidence level are modified in an optimal fashion to incorporate the a priori information that the parameter being estimated is non-negative. A paradox with different confidence intervals for the same confidence level is clarified. The "lossy compression" nature of the device of confidence intervals is discussed and a "lossless" option to present results is pointed out.

The present Letter is an expanded version of an excerpt removed from ref. [1] in the last minute (whence some numbering glitches in the first posted version of ref. [1]). The excerpt dealt with a variation on the theme of ref. [1] but was found to contain a paradox. Although not a defect of logic or mathematics, the paradox deserved to be clarified, whereas the primary result of ref. [1] was felt to deserve an undistracted presentation. The purpose of the present Letter is to give a proper treatment to the result and the paradox from that excerpt. The notations of ref. [1] are used without further explanations; all numbered references are to the second posting of ref. [1].

## 1. Modifying the upper bound

Ref. [1] started from a conventional estimator $\hat{\theta}$ for the parameter $\theta$ and redefined it so as to take into account the a priori inequality $\theta \geq 0$. Then for the redefined estimator $\tilde{\theta} = \max(\hat{\theta}, 0)$ (eq. (9) of [1]), a conventional confidence belt was constructed in a more or less straightforward fashion. The treatment of the $\delta$-functional contribution to the probability distribution of $\tilde{\theta}$ was simplified via an observation that reduced the problem to constructing a confidence belt for the unmodified estimator $\hat{\theta}$ in such a way that the resulting belt satisfy an additional condition (sec. 3 of [1]). The construction was acccomplished using the trick of so-called horizontal deformations (sec. 2 of [1]), with the result represented by Fig. 6 of [1].

Ref. [1] modified the standard symmetric confidence belt, which corresponds to the option $\alpha = \alpha' = (1-\beta)/2$ in terms of Fig. 1 of [1]. A natural variation on the same theme is to accomplish a similar modification for the asymmetric case $\alpha' = 0$, $\beta = 1 - \alpha$ that corresponds to an upper bound for the confidence level $\beta$:

$$\mathsf{P}\left(\theta < l_{1-\beta}(\hat{\theta})\right) = \beta \tag{1}$$

This option is useful when one is trying to measure a positive signal whereas the statistical accuracy may not be high enough to establish a non-zero signal with a high confidence. Then one would like to establish as tight an upper bound as possible.

We are going to modify the confidence belt (1) to accomodate the a priori inequality $\theta \geq 0$. The required geometrical infrastructure is provided by Fig. 1 that differs from Fig. 2 of [1] by adding a few more intersection points (the intersection points MBDN on the horizontal line LG).

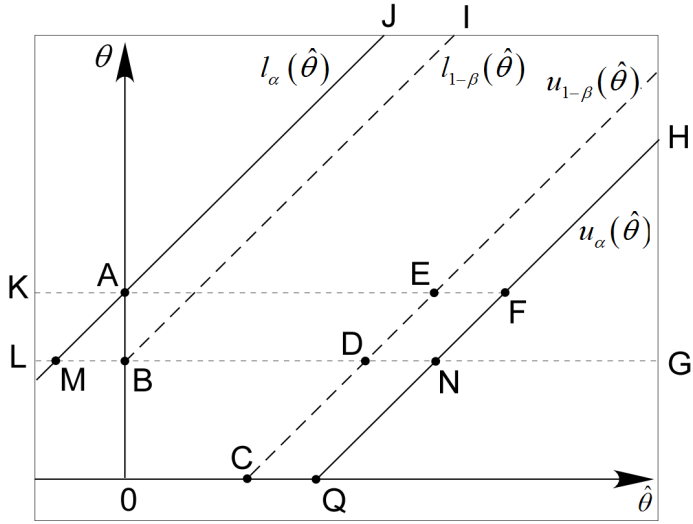

**Fig. 1.** The pairs of solid and dashed sloping lines delimit symmetric confidence belts for the confidence limits $\beta = 1 - 2\alpha$ and $\tilde\beta = 1 - 2(1-\beta) = 1 - 4\alpha$. The functions that correspond to the lines are shown in the figure. A and B are intersections with the vertical axis of the lines $\theta = l_\alpha(\hat\theta)$ and $\theta = l_{1-\beta}(\hat\theta)$. Points A and B determine the horizontal lines KF and LG along with further intersection points.

Only the points M, B, C, D, N will play a role in what follows; the other points are shown to establish a connection with Fig. 2 of [1].

The number $\theta_B$ is the vertical position of the intersection point B (and of M, D, and N):

$$\theta_B = l_{1-\beta}(0) \qquad (2)$$

The numbers $\theta_C < \theta_D$ are the horizontal positions of the points C and D:

$$\theta_C = u_{1-\beta}(0), \quad \theta_D = u_{1-\beta}(\theta_B) \qquad (3)$$

The unmodified bound (1) corresponds to confidence intervals (level $\beta$) that start on the upper dashed line BI and stretch down to infinity.

To obtain a modified version of the bound (1), one starts from an allowed confidence belt $[u(\hat\theta), l(\hat\theta)]$ shown with the fat lines in Fig. 2.

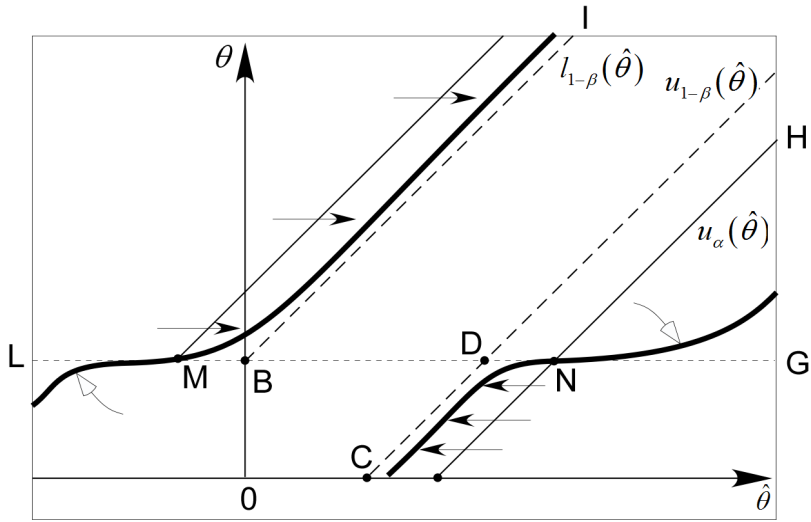

**Fig. 2.** The two fat curves delimit an allowed confidence belt for the confidence level $\beta$. The fat lines are hinged at the points M and N. Black arrows show allowed horizontal deformations.
White arrows show the resulting straightening of the corresponding segments.

Then one performs the horizontal deformations of $u$ and $l$ as shown by the black arrows (detailed explanations of the trick are given in sec. 2 of [1]). Then the lower segment of $u$ below point N is pressed to the straight segment CD, whereas the upper segment of $l$ above point M is pressed to BI.
The resulting effective deformations on the other side are shown by white arrows.
The confidence belt thus obtained is shown in Fig. 3.

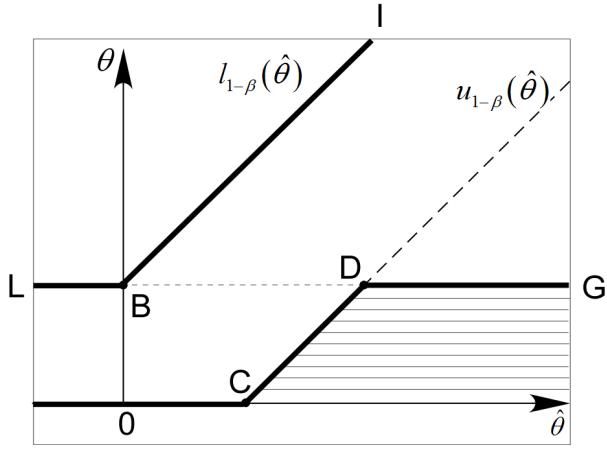

**Fig. 3.** The confidence belt obtained from the unmodified upper bound for the confidence level $\beta$ by taking into account the a priori information $\theta \geq 0$.

The region under CDG is a pure gain from the a priori information.

The analytical description is as follows (we are talking about the confidence level $\beta$):

— For $\hat{\theta} \geq \theta_D$, the confidence interval is $\left[\theta_B, l_{1-\beta}(\hat{\theta})\right]$, i.e. the upper bound is the same as in the unmodified case, eq. (1), but restricted from below at $\theta_B$. The region under CDG is exactly the gain from the a priori information.

— For $\theta_C \leq \hat{\theta} \leq \theta_D$, the confidence interval is $\left[u_{1-\beta}(\hat{\theta}), l_{1-\beta}(\hat{\theta})\right]$, i.e. exactly the unmodified symmetric confidence interval for the confidence level $\tilde{\beta} = 1 - 2(1-\beta) = 1 - 4\alpha$.

— For $0 \leq \hat{\theta} \leq \theta_C$, the confidence interval is $\left[0, l_{1-\beta}(\hat{\theta})\right]$, i.e. the unmodified bound (1) restricted from below by the physical boundary.

— Lastly, for $\hat{\theta} \leq 0$, the confidence interval is fixed as $[0, \theta_B]$.

The noteworthy properties of this confidence belt are as follows:

♥ The estimate is robust for non-physical values of the estimator, i.e. for $\hat{\theta} < 0$.

♥ The interval's upper bound for physical values of $\hat{\theta}$ is the same as in the unmodified case (15) and is the lowest possible one at the confidence level $\beta$.

♥ The interval's lower bound breaks off zero at the earliest point possible for the given confidence level ($\theta_C$), and the lower bound is maximal possible for this confidence level in the interval $\theta_C \leq \hat{\theta} \leq \theta_D$.

♥ Neither complicated algorithms nor tables are required on top of the standard routines to compute the confidence interval for the confidence level $\tilde{\beta} = 1 - 2(1-\beta) = 1 - 4\alpha$.

## 2. The paradox

If one compares the confidence belt of Fig. 3 above with that of Fig. 6 of [1], one sees that for all $\hat{\theta} < \theta_D$, the lower bounds of the two intervals coincide whereas the upper bound of the former is strictly below that of the latter — for the same confidence level $\beta$. In other words, for a range of values of $\hat{\theta}$, the belt of Fig. 3 yields strictly tighter confidence intervals than that of Fig. 6 of [1] — for the same confidence level $\beta$.

This phenomenon is unrelated to the a priori inequality $\theta \geq 0$ as it can be observed already with conventional confidence belts, as shown in Fig. 4.

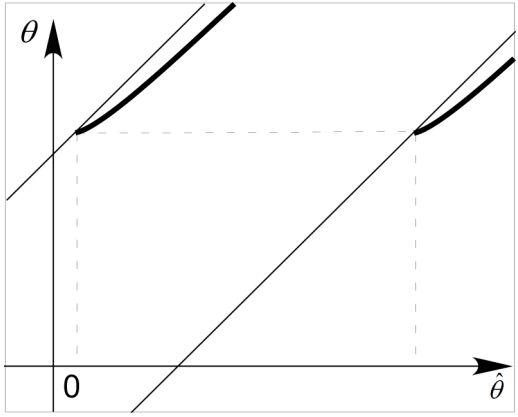

**Fig. 4.** Fat lines show a horizontal deformation of the confidence belt that does not change the confidence level $\beta$, but results in strictly tighter confidence intervals for the values of the estimator within the segment delimited by the two dashed vertical lines.

The example was pointed out by A.V. Lokhov.

To understand the apparent paradox one has to recall the meaning of confidence intervals, namely, that the probability for the random interval $[u(\hat{\theta}), l(\hat{\theta})]$ to cover the unknown $\theta$ is $\beta$. This is formally expressed by the following expression (cf. eq. (4) of [1]):

$$\beta = \mathsf{P}\left(u(\hat{\theta}) < \theta < l(\hat{\theta})\right) \equiv \mathsf{P}\left(L(\theta) < \hat{\theta} < U(\theta)\right) \equiv \int_{L(\theta)}^{U(\theta)} d\hat{\theta}\, d_\theta(\hat{\theta}) \qquad (4)$$

This is an integral relation, and it only guarantees that for any fixed value of the unknown parameter $\theta$ the integration region $[L(\theta), U(\theta)]$ cannot become tighter if the confidence level is to remain fixed at the value $\beta$. Eq. (4) guarantees exactly nothing for any specific value of the estimator.

This is similar to how positivity of an integral does not imply positivity of the integrand at any point of the integration region. So does eq. (4) not imply anything in regard of the pair $[u(\hat{\theta}), l(\hat{\theta})]$ at any value of $\hat{\theta}$.

In other words, the apparent paradox contradicts the naive expectations about how confidence intervals work — but there is neither mathematical nor logical contradiction.

## 3. Discussion

Generally speaking, a complete and unambiguous way to present results of an experiment is to provide, along with the measured value of $\hat{\theta}$, not a confidence interval but the complete density $d_\theta(\hat{\theta})$. Such a density summarizes all the information about the experiment in regard of measurement of $\beta$, and the measured value of $\hat{\theta}$ represents the actual outcome of the experiment. At this level of reasoning, the problem of a priori information does not occur: the a priori inequality $\theta \geq 0$ is part of the definition of $d_\theta(\hat{\theta})$. Note that it is fairly easy to describe a function of two real arguments, especially that $d_\theta(\hat{\theta})$ usually has a simple qualitative behavior.

Anything beyond that is a device similar to what is called *lossy compression* in the computer industry. Such a device is based on a decision that involves extrinsic considerations and that is guided but not fully determined by the mathematical statistics proper. Only after the rules to calculate wins are explicitly stated (by the casino, QA department, or the Nobel Committee) can one meaningfully choose an optimal confidence interval. Compressed descriptions of complex data are ubiquitous (approximations etc.); their purpose is to represent the salient features of the data, and there normally *is* an arbitrariness involved.

The loss of information resulting from such a compression may be vanishing in special cases, e.g. for a Poisson distribution with $\mu = \sigma = \theta$ or for a normal distribution $N(\theta, \sigma)$ with $\sigma$ independent of $\theta$. For more complicated $d_\theta(\hat{\theta})$, however, quoting one confidence interval for a given confidence limit may imply a loss of information about $d_\theta(\hat{\theta})$. In any case, there is no harm in quoting a few different confidence intervals even for the same confidence level, e.g. a modified symmetric interval as defined in ref. [1], a Feldman-Cousins interval [2], and a modified upper bound as defined in Fig. 3 above, for the same unmodified estimator $\hat{\theta}$, provided one clearly indicates which is which.

Speaking of presenting several different confidence intevals simultaneously, there is actually a way to do so in a visual fashion. It falls short of, but is simpler than presenting the full $d_\theta(\hat{\theta})$.
On the other hand, it is not plagued by arbitrariness as is the case with confidence intervals, and confidence intervals for any prescription can be read off the graph defined as follows. Define two functions:

$$A(\theta) = \int_{-\infty}^{\hat{\theta}} d\hat{\theta}'\, d_\theta(\hat{\theta}'), \quad A'(\theta) = \int_{\hat{\theta}}^{+\infty} d\hat{\theta}'\, d_\theta(\hat{\theta}') \tag{5}$$

Then the quantities $l_\alpha(\hat{\theta})$ and $u_{\alpha'}(\hat{\theta})$ that for various $\alpha$ and $\alpha'$ represent the boundaries of various confidence intervals (cf. Fig. 1), are solutions of the following equations:

$$A(l_\alpha(\hat{\theta})) = \alpha, \quad A'(u_{\alpha'}(\hat{\theta})) = \alpha' \tag{6}$$

These equations can be easily solved in a graphical fashion using a plot of $\max(A, A')$ as a function of $\theta$ for the measured $\hat{\theta}$ using horizontal lines for various confidence levels. There will be a peak near $\theta = \hat{\theta}$ with $A = A' = \frac{1}{2}$. As with $d_\theta(\hat{\theta})$, no special measures are necessary to take into account the a priori bound $\theta \geq 0$: the plot is simply limited to physical values of $\theta$. From such a plot any confidence interval — (a)symmetric, (un)modified, etc. — for any confidence level can be deduced.

## 4. Summary

The modified upper bound prescription of Fig. 3 correctly takes into account the a priori lower bound $\theta \geq 0$. It complements the modified confidence belt prescription of ref. [1]. It is more suitable for experiments where one primarily aims to establish an upper bound rather then to measure a non-zero parameter; however, an earliest resolution from zero is still guaranteed.

The apparent paradox is that different confidence belt prescriptions can yield embedded confidence intervals for the same confidence level, one strictly tighter than another. The paradox is explained by the statistical nature of the statement that a given confidence interval covers the unknown exact value with a given probability (sec. 2).

As to how measurements results should be presented, a complete and unambiguous presentation would consist of the measured value of $\hat{\theta}$ as well as the probability density $d_\theta(\hat{\theta})$ in some form. A less complete — but still unambiguous — presentation involves the plot of $A(\theta)$ and $A'(\theta)$ computed for the measured $\hat{\theta}$, with horizontal confidence level lines as explained in sec. 3.
A third way, even less complete but sufficiently informative in many cases, is to employ the prescriptions of this Letter or/and ref. [1].

However, if $d_\theta(\hat{\theta})$ is sufficiently close to, say, the normal or Poisson distribution, then it is sufficient to present its variance besides stating that it is a normal or Poisson distribution. This would provide a complete information about $d_\theta(\hat{\theta})$. Then confidence intervals constructed using different prescriptions (the ones of the present Letter or refs. [1] and [2]) are essentially equivalent ways to present the same information about the measured value of $\hat{\theta}$ and the distribution $d_\theta(\hat{\theta})$. One should, of course, avoid handmade patchwork constructions (for lack of an adequate translation of the Russian adjective *рукосуйские*) like the one criticised by Feldman and Cousins [2].

*Acknowledgements.* A.S. Barabash offered encouragement when confronted with the first sketch of Fig. 3. A.V. Lokhov pointed out Fig. 4. The stimuli for developing the lossy compression argument of sec. 3 were V.Z. Nozik's advice to not hesitate to quote results obtained with different estimators as long as the methods employed are clearly indicated, and A.A. Nozik's making fun of the firmness of some experimentalists' belief in the only one "correct" confidence interval.

Thanks are also due to the members of the Troitsk ν-mass experiment for providing a stimulating context for this work.